\documentclass[doublecol]{epl2} 

\usepackage{amsfonts, amsmath, amssymb, bbold, bigints, color, epstopdf, float, graphicx, lineno, makeidx, rotating, setspace, subfigure, url}
\usepackage[bookmarks, breaklinks, colorlinks]{hyperref}
\hypersetup{citecolor = blue, filecolor = black, linkcolor = red, urlcolor = blue}
\hypersetup{pdfauthor = {Chiranjit Mitra}}

\newcommand{\DSF}{\textsc{DSF}}
\newcommand{\PSF}{\textsc{PSF}}
\newcommand{\MSF}{\textsc{MSF}}

\title{Rewiring hierarchical scale-free networks: Influence on synchronizability and topology}
\shorttitle{Rewiring hierarchical scale-free networks: Influence on synchronizability and topology} 

\author{Chiranjit Mitra\inst{1, 2} \and J\"{u}rgen Kurths\inst{1, 2} \and Reik V. Donner\inst{1}}
\shortauthor{C. Mitra \etal}

\institute{
  \inst{1} Research Domain IV - Transdisciplinary Concepts \& Methods, Potsdam Institute for Climate Impact Research, 14473 Potsdam, Germany\\
  \inst{2} Department of Physics, Humboldt University of Berlin, 12489 Berlin, Germany
}


\abstract{Many real-world complex networks simultaneously exhibit topological features of scale-free behaviour and hierarchical organization. In this regard, deterministic scale-free [A.-L. Barab{\'a}si \etal, Physica A, 299, 3 (2001)] and pseudofractal scale-free [S. N. Dorogovtsev \etal, Phy. Rev. E, 65, 6 (2002)] networks constitute notable models which simultaneously incorporate the aforementioned properties. The rules governing the formation of such networks are completely deterministic. However, real-world networks are presumably neither completely deterministic, nor perfectly hierarchical. Therefore, we suggest here perfectly hierarchical scale-free networks with randomly rewired edges as better representatives of practical networked systems. In particular, we preserve the scale-free degree distribution of the various deterministic networks but successively relax the hierarchical structure while rewiring them. We utilize the framework of master stability function in investigating the synchronizability of dynamical systems coupled on such rewired networks. Interestingly, this reveals that the process of rewiring is capable of significantly enhancing, as well as, deteriorating the synchronizability of the resulting networks. We investigate the influence of rewiring edges on the topological properties of the rewired networks and, in turn, their relation to the synchronizability of the respective topologies. Finally, we compare the synchronizability of deterministic scale-free and pseudofractcal scale-free networks with that of random scale-free networks (generated using the classical Barab\'{a}si-Albert model of growth and preferential attachment) and find that the latter ones promote synchronizability better than their deterministic counterparts.}

\begin{document}

\maketitle


\section{\label{sec:Introduction}Introduction}

Complex systems involving large collections of dynamical elements interacting with each other on complex networks are abundant across several disciplines of sciences and engineering~\cite{albert2002statistical, dorogovtsev2002evolution, boccaletti2006complex, newman2010networks}. This has generated a consolidated effort towards unveiling structural properties of manifold real-world networked systems and uncovering fundamental principles governing their organization~\cite{newman2003structure}. A significant milestone amid such explorations was the exposition of the small-world behaviour of diverse real networks, characterized by a small average path length between nodes and a high clustering coefficient~\cite{watts1998collective}. Further, the interplay between topological properties of complex networked systems and the collective dynamics exhibited by them has been simultaneously investigated, particularly with reference to the phenomenon of synchronization~\cite{pikovsky2003synchronization, arenas2008synchronization, newman2010networks}.

Synchronization is among the most relevant emergent behaviours in complex networks of dynamical systems and is often critical to their functionality~\cite{pikovsky2003synchronization, arenas2008synchronization, menck2014dead, mitra2014dynamical, mitra2015integrative, mitra2017multiple, mitra2017recovery}. As a result, there has been a persistent drive towards unravelling the influence of topological features of networks on their ability to synchronize, often with the objective of designing topologies for better synchronizability~\cite{motter2005enhancing, motter2005network, donetti2005entangled, nishikawa2006synchronization, nishikawa2006maximum, yin2006decoupling, duan2007complex, motter2007bounding, gu2009altering, nishikawa2010network}. In this regard, small-world networks have been particularly known to facilitate synchronization of dynamical systems coupled on them~\cite{lago2000fast, gade2000synchronous, hong2002synchronization, wang2002synchronizationsmall, barahona2002synchronization}. Besides the small-world property, real-world networks often exhibit two other remarkable generic features, namely, scale-free behaviour~\cite{barabasi1999emergence} and hierarchical structure~\cite{ravasz2002hierarchical, clauset2008hierarchical}.

Scale-free behaviour is characterized by the probability $P \left( k \right)$ that a randomly selected node has exactly $k$ links decaying as a power law $\left( P \left( k \right) \sim k^{- \gamma} \right)$ and appears in good approximation in diverse real networked systems such as the internet~\cite{faloutsos1999power}, the world wide web~\cite{barabasi1999emergence}, networks of metabolic reactions~\cite{jeong2000large}, protein interaction networks~\cite{jeong2001lethality}, the web of Hollywood actors linked by movies~\cite{albert2000topology}, social networks such as the web of human sexual contacts~\cite{liljeros2001web}, etc. In this context, the Barab\'{a}si-Albert (BA) model~\cite{barabasi1999emergence} has been suggested for realizing random scale-free networks with growth and preferential attachment, where an incoming node is more likely to get randomly linked to an existing node with higher connectivity.

Also, manifold real-world systems such as metabolic networks in the cell~\cite{ravasz2002hierarchical}, ecological niches in food webs~\cite{clauset2008hierarchical}, the scientific collaboration network~\cite{shen2009detect}, corporate and governmental organizations~\cite{yu2006genomic}, etc.\ exhibit hierarchical organization where small groups of nodes organize in a stratified manner into larger groups, over multiple scales. This definition of hierarchical structure, also used throughout this letter, relates to that proposed by Clauset \etal~\cite{clauset2008hierarchical}.

Naturally, collective dynamics on scale-free~\cite{jost2001spectral, wang2002synchronizationscale, wang2002complex, lind2004coherence} and hierarchical topologies~\cite{arenas2006synchronization, diaz2008dynamics, skardal2012hierarchical, mitra2017multiple, mitra2017recovery} have been investigated intensively, but mostly separately, leaving sufficient room for further explorations concerning synchronization in networks simultaneously exhibiting the two topological properties mentioned above. Notably, the coexistence of the generic feature of scale-free topology along with a hierarchical organization in many networks in nature and society is immensely intriguing~\cite{ravasz2003hierarchical}. Examples in this direction constitute the internet at the domain level, the world wide web of documents, the actor network, the semantic web viewed as a network of words, biochemical networks in the cell, etc.~\cite{ravasz2002hierarchical, ravasz2003hierarchical}.

\begin{figure}
\begin{center}
\subfigure[]
{
\includegraphics[height=2.5cm, width=5.0cm]{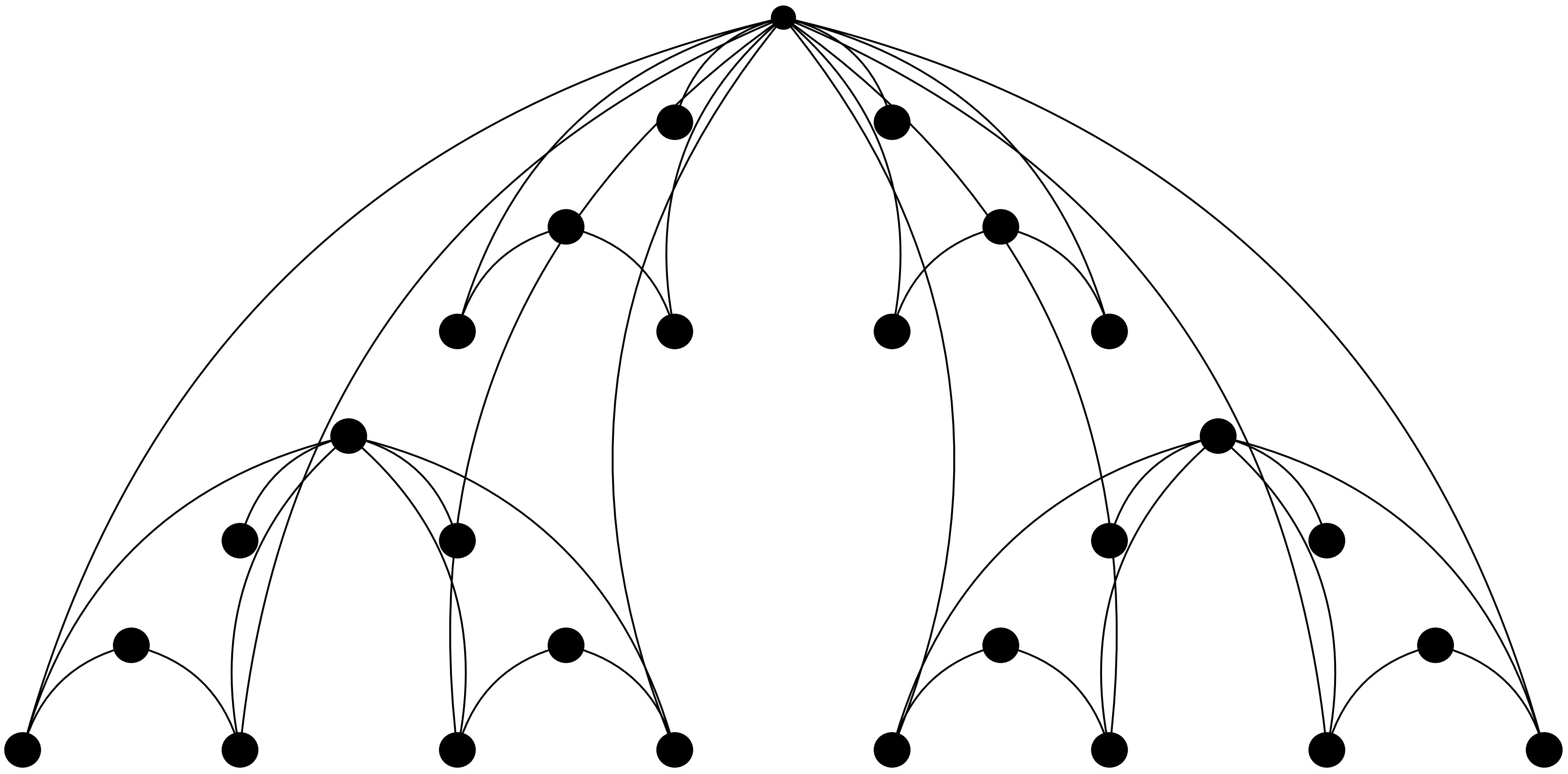}
}
\subfigure[]
{
\includegraphics[height=2.5cm, width=3.0cm]{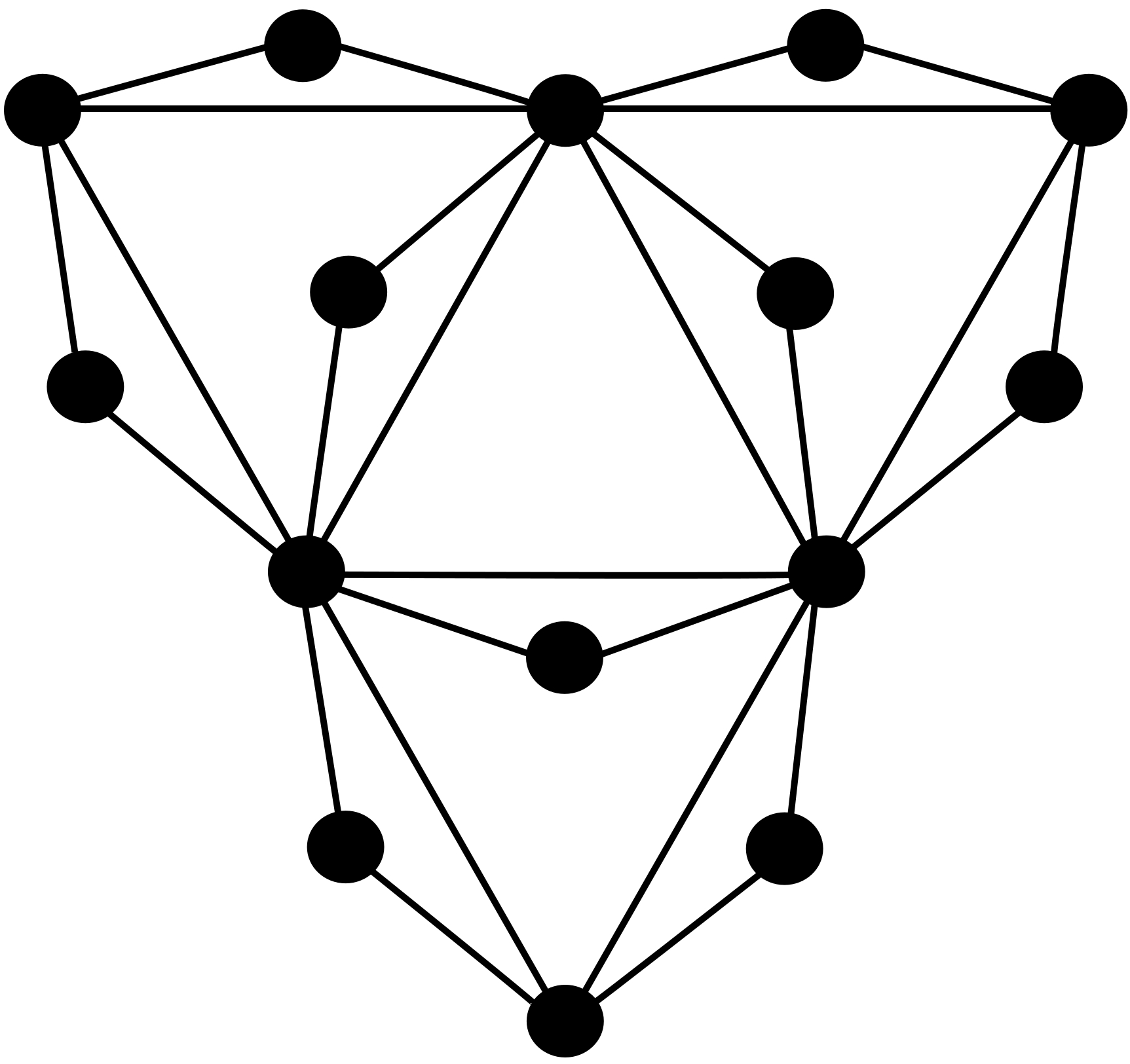}
}
\caption{\label{fig:Figure_1}Topology of the (a) deterministic and (b) pseudofractal scale-free networks developed over 2 generations.}
\end{center}
\end{figure}


\subsection{\label{sec:Network_Construction}Network Construction}

\begin{figure}
\begin{center}
\subfigure[]
{
\includegraphics[height=2.5cm, width=2.5cm]{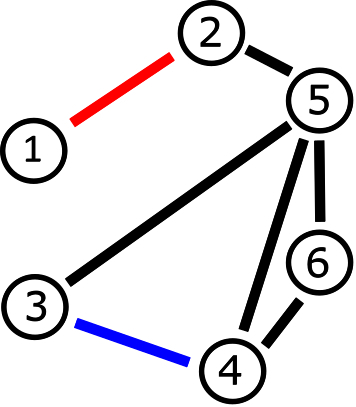}
}
\hspace{0.1cm}
\subfigure[]
{
\includegraphics[height=2.5cm, width=2.5cm]{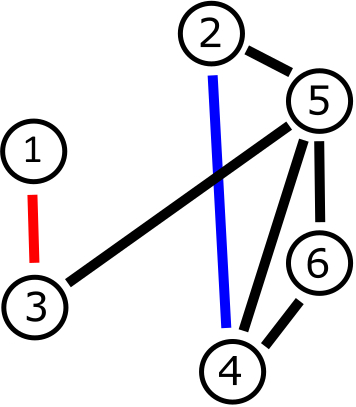}
}
\hspace{0.1cm}
\subfigure[]
{
\includegraphics[height=2.5cm, width=2.5cm]{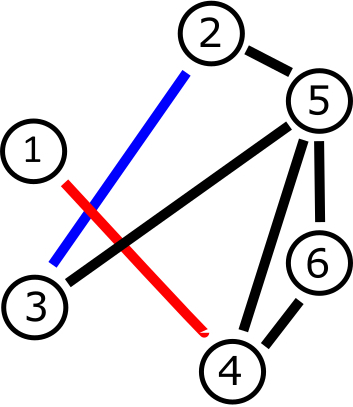}
}
\caption{\label{fig:Figure_2}(Color online) (a) We randomly select two (distinct) edges of the network with the first edge (red) connecting nodes numbered 1 and 2 and the second edge (blue) connecting nodes numbered 3 and 4. We rewire (b) the first edge to connect nodes 1 and 3 and the second edge to connect nodes 2 and 4 (provided there does not already exist an edge between nodes 1 and 3 or between 2 and 4). Otherwise, we rewire (c) the first edge to connect nodes 1 and 4 and the second edge to connect nodes 2 and 3 (provided there does not already exist edges between the respective nodes as well). If the aforementioned steps fail, we choose a new pair of edges to rewire. Clearly, we preserve the scale-free degree distribution of the deterministic networks we start with, but successively loose the hierarchical structure while rewiring them. Also, note that we allow for a multiple selection of the same edge in subsequent rewiring steps.}
\end{center}
\end{figure}

Notable instances among models simultaneously incorporating the prominent topological features of scale-free behaviour and hierarchical organization under one roof are the deterministic scale-free (\DSF{})~\cite{barabasi2001deterministic}, pseudofractal scale-free (\PSF{})~\cite{dorogovtsev2002pseudofractal}, Apollonian~\cite{andrade2005apollonian} and the hierarchical network model~\cite{ravasz2003hierarchical}. We specifically study \DSF{} and \PSF{} networks in this letter, the topology of them developed over 2 generations is illustrated in Fig.~\ref{fig:Figure_1}(a, b). Evidently, these models are completely deterministic, leading to a perfectly hierarchical assembly of the associated networks. However, it is most natural to assume that real-world topologies are neither completely deterministic, nor perfectly hierarchical. Thus, a realistic model of practical networked systems should feature an aspect of randomness, besides simultaneously manifesting not far from scale-free and hierarchical design. Henceforth, as a preliminary solution to this problem, we suggest in the following perfectly hierarchical networks (generated by the deterministic rules of the aforementioned models) with randomly rewired links as better representatives of associated connected architectures in the real-world. The mechanism used throughout this letter for rewiring edges, while preserving the (scale-free) degree distribution of the otherwise perfectly hierarchical networks, is illustrated in Fig.~\ref{fig:Figure_2}.

The desired operational state in complex networks is often associated with the synchronized motion of its dynamical components~\cite{pikovsky2003synchronization}. In this work, we investigate the synchronizability of the proposed network models using the master stability function (\MSF{}) framework~\cite{pecora1998master}. We recall that real-world topologies exhibiting the small-world property are known to facilitate network synchronization~\cite{nishikawa2003heterogeneity, menck2013basin}, as well as, to be more robust to random perturbations~\cite{menck2013basin}. In this regard, the classical network model of Watts and Strogatz~\cite{watts1998collective} is particularly notable for capturing the small-world property. In strong analogy with the present work, the Watts-Strogatz model generates graphs by randomly rewiring completely regular architectures (ring lattices), thus interpolating between absolutely regular and random graphs with the small-world property appearing for intermediate rewiring. However, \MSF{}-based~\cite{pecora1998master} measurements of synchronizability of the Watts-Strogatz model~\cite{watts1998collective} surprisingly do not reveal exclusive features in the small-world regime~\cite{hong2002synchronization}. In such networks, synchronizability is only enhanced for an initial increase of the number of rewired edges, which then saturates afterwards as further links are rewired. In fact, synchronizability of the rewired networks (for a given number of rewired edges) are not much different from one another. On the other hand, networks resulting from rewiring hierarchical scale-free networks considered here exhibit both significantly enhanced, as well as, deteriorated synchronizability (compared to that of their completely deterministic counterparts).


\section{\label{sec:Methods}Methods}

In the following, we briefly review the framework of MSF~\cite{pecora1998master} and the traditional quantifier of synchronizability of a network, prior to its application to the aforementioned network models. Subsequently, we discuss a few key characteristics of network topology and the relationships between them with the synchronizability of the networks will be studied in this letter.


\subsection{\label{sec:Synchronizability}Synchronizability}

Consider a network of $N$ identical oscillators where the isolated dynamics of the $i\textsuperscript{th}$ oscillator is described by
\begin{equation} \label{eq:DE_Individual}
\dot{\mathbf{x}}^{i} = \mathbf{F} \left( \mathbf{x}^{i} \right);\, \mathbf{x}^{i} \in \mathbb{R}^{d},\, i = 1,\, 2,\, \ldots,\, N,
\end{equation}
and coupling is established via an output function $\mathbf{H}:\, \mathbb{R}^{d}\, \rightarrow\, \mathbb{R}^{d}$ (identical for all $i$). The topology of interactions is captured by the adjacency matrix $\mathbf{A}$, where $A_{ij} = 1$ if nodes $i$ and $j \left( \neq i \right)$ are connected while $A_{ij} = 0$ otherwise. The dynamical equations of the networked system read
\begin{equation} \label{eq:DE_Network}
\begin{split}
\dot{\mathbf{x}}^{i} & = \mathbf{F} \left( \mathbf{x}^{i} \right) + \epsilon \sum\limits_{j = 1}^{N} A_{ij} \left[ \mathbf{H} \left( \mathbf{x}^{j} \right) - \mathbf{H} \left( \mathbf{x}^{i} \right) \right]\\
& = \mathbf{F} \left( \mathbf{x}^{i} \right) - \epsilon \sum\limits_{j = 1}^{N} L_{ij} \mathbf{H} \left( \mathbf{x}^{j} \right)
\end{split}
\end{equation}
where $\epsilon$ represents the overall coupling strength and $\mathbf{L}$ is the graph Laplacian such that $L_{ij} = - A_{ij}$ if $i \neq j$ and $L_{ii} = \sum\limits_{j = 1}^{N} A_{ij} = k_{i}$ is the degree of node $i$. Since the Laplacian matrix $\mathbf{L}$ is symmetric, its eigenvalue spectrum $\left( \lambda_{1},\, \lambda_{2},\, \ldots,\, \lambda_{N} \right)$ is real and ordered as $0 = \lambda_{1} < \lambda_{2} \le \ldots \le \lambda_{N}$, assuming the network is connected. Further, $\mathbf{L}$ has zero row sum by definition, guaranteeing the existence of a completely synchronized state, $\mathbf{x}^{1} \left( t \right) = \mathbf{x}^{2} \left( t \right) = \ldots = \mathbf{x}^{N} \left( t \right) = \mathbf{s} \left( t \right)$ as a solution of Eq.~(\ref{eq:DE_Network}). Starting from heterogeneous initial conditions, the oscillators (asymptotically) approach (and thus evolve on) the synchronization manifold $\mathbf{s} \left( t \right)$ corresponding to the solution of the uncoupled dynamics of the individual oscillators in Eq.~(\ref{eq:DE_Individual}) $\left( \dot{\mathbf{s}} = \mathbf{F} \left( \mathbf{s} \right) \right)$.

The local stability of the completely synchronized state determined by the framework of MSF~\cite{pecora1998master} relates the \emph{synchronizability} of a network to the \emph{eigenratio} $R \equiv \frac{\lambda_{N}}{\lambda_{2}}$. Irrespective of $\mathbf{F}$ and $\mathbf{H}$ (Eq.~(\ref{eq:DE_Network})), this condition has been extensively used to characterize the synchronizability of a network such that the lower the value of $R$, the more synchronizable the network and vice versa~\cite{barahona2002synchronization, nishikawa2003heterogeneity, motter2005enhancing, andrade2005apollonian, motter2005network, donetti2005entangled, boccaletti2006complex, nishikawa2006synchronization, nishikawa2006maximum, yin2006decoupling, duan2007complex, motter2007bounding, arenas2008synchronization, gu2009altering, nishikawa2010network, rad2008efficient, dadashi2010rewiring, jalili2013enhancing}.


\subsection{\label{sec:Network_Properties}Network Properties}

We utilize the above framework in exploring the synchronizability of the aforementioned network models (Fig.~\ref{fig:Figure_1}) after stochastically rewiring their edges. Further, we investigate the influence of rewiring on the topological properties of the resulting networks and in turn, their relation to the synchronizability of the associated topologies. For that purpose, we now briefly describe the topological properties of average path length, maximum betweenness centrality, average local clustering coefficient and global clustering coefficient (transitivity) of a network.

The \emph{average path length} $\mathcal{L}$ of a network with $N$ nodes is defined as the mean value of the shortest path length between all possible pairs of nodes~\cite{newman2010networks}. Thus, $\mathcal{L} = \frac{1}{N \left( N - 1 \right)} \sum\limits_{i \neq j} \ell \left( i, j \right)$ where $\ell \left( i, j \right)$ is the length of the shortest path between nodes $i$ and $j$~\cite{newman2010networks}. Intuitively, a smaller average path length of a network should facilitate efficient communication between oscillators, culminating in improved synchronizability of the overall system~\cite{hong2002synchronization}.

The \emph{betweenness centrality} $bc_{i}$ of a node $i$ is related to the fraction of shortest paths between all pairs of nodes that pass through that node~\cite{newman2010networks}. For an $N$-node network, the betweenness centrality of each node may further be normalized by dividing it by the number of node pairs $\left( \text{i.e.},\ {N \choose 2} \right)$, resulting in values between 0 and 1. Thus, $bc_{i} = \frac{2}{N \left( N - 1 \right)} \sum\limits_{j \neq k \neq i} \frac{\sigma_{j, k}^{i}}{\sigma_{j, k}}$, where $\sigma_{j, k}$ is the total number of shortest paths from node $j$ to node $k$ and $\sigma_{j, k}^{i}$ is the number of such shortest paths which pass through node $i$~\cite{newman2010networks}. We study here the maximum betweenness centrality values $bc_{max}$ of all nodes of a network realization, which have been argued to be inversely related to synchronizability~\cite{hong2004factors}.

The \emph{local clustering coefficient} $\mathcal{C}_{i}^{L}$ relates to the probability of the existence of an edge between two randomly selected neighbours of node $i$~\cite{newman2010networks}. $\mathcal{C}_{i}^{L}$ is defined as the ratio between the number of links between nodes within the neighbourhood of node $i$ and the number of links that could possibly exist between its neighbours~\cite{newman2010networks}. Thus, $\mathcal{C}_{i}^{L} = \frac{2}{k_{i} \left( k_{i} - 1 \right)} N_{i}^{\Delta}$ where $N_{i}^{\Delta}$ is the total number of closed triangles including node $i$ (with degree $k_{i}$), which is bounded by the maximum possible value of $\frac{k_{i} \left( k_{i} - 1 \right)}{2}$~\cite{newman2010networks}. The \emph{average local clustering coefficient} $\mathcal{C}^{L}$ of the network is then given by the mean of the local clustering coefficients of all nodes of the network, i.e., $\mathcal{C}^{L} = \frac{1}{N} \sum\limits_{i = 1}^{N} \mathcal{C}_{i}^{L}$. Likewise, the \emph{global clustering coefficient} $\mathcal{C}^{G}$ of a network (often also called network \emph{transitivity}~\cite{newman2010networks, barrat2000properties}) is related to the mean probability that two nodes with a common neighbour are themselves neighbours~\cite{newman2010networks}. $\mathcal{C}^{G}$ is defined as the fraction of the total number of triplets in the network that are closed, i.e, $\mathcal{C}^{G} = \frac{\text{(number of closed triplets)}}{\text{(total number of triplets)}}$~\cite{newman2010networks}. In this case, a triplet means three vertices $i$, $j$ and $k$ with edges $\left( i,\, j \right)$ and $\left( j,\, k \right)$, while the edge $\left( i,\, k \right)$ may be present or not. To avoid terminological confusion, we emphasize that the average local clustering coefficient $\mathcal{C}^{L}$ (as defined in this letter) is often referred to as the global clustering coefficient (e.g., as in Ref.~\cite{watts1998collective}). Larger clustering coefficients are generally associated with a reduced synchronizability of small-world and scale-free networks~\cite{arenas2008synchronization}.


\section{\label{sec:Results}Results}

\begin{figure}
\begin{center}
\subfigure[]
{
\includegraphics[height=4.75cm, width=8.5cm]{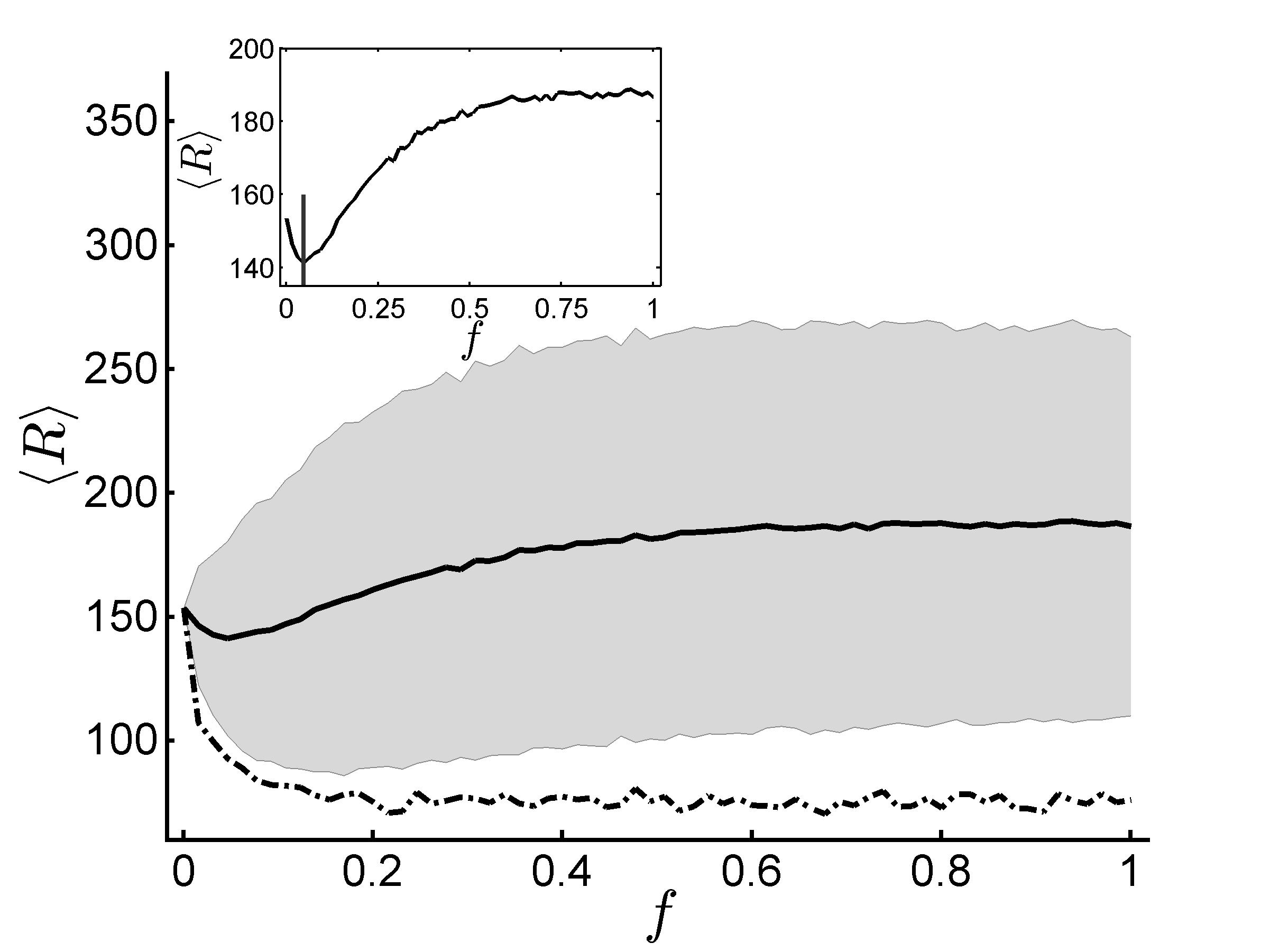}
}
\\
\subfigure[]
{
\includegraphics[height=4.75cm, width=8.5cm]{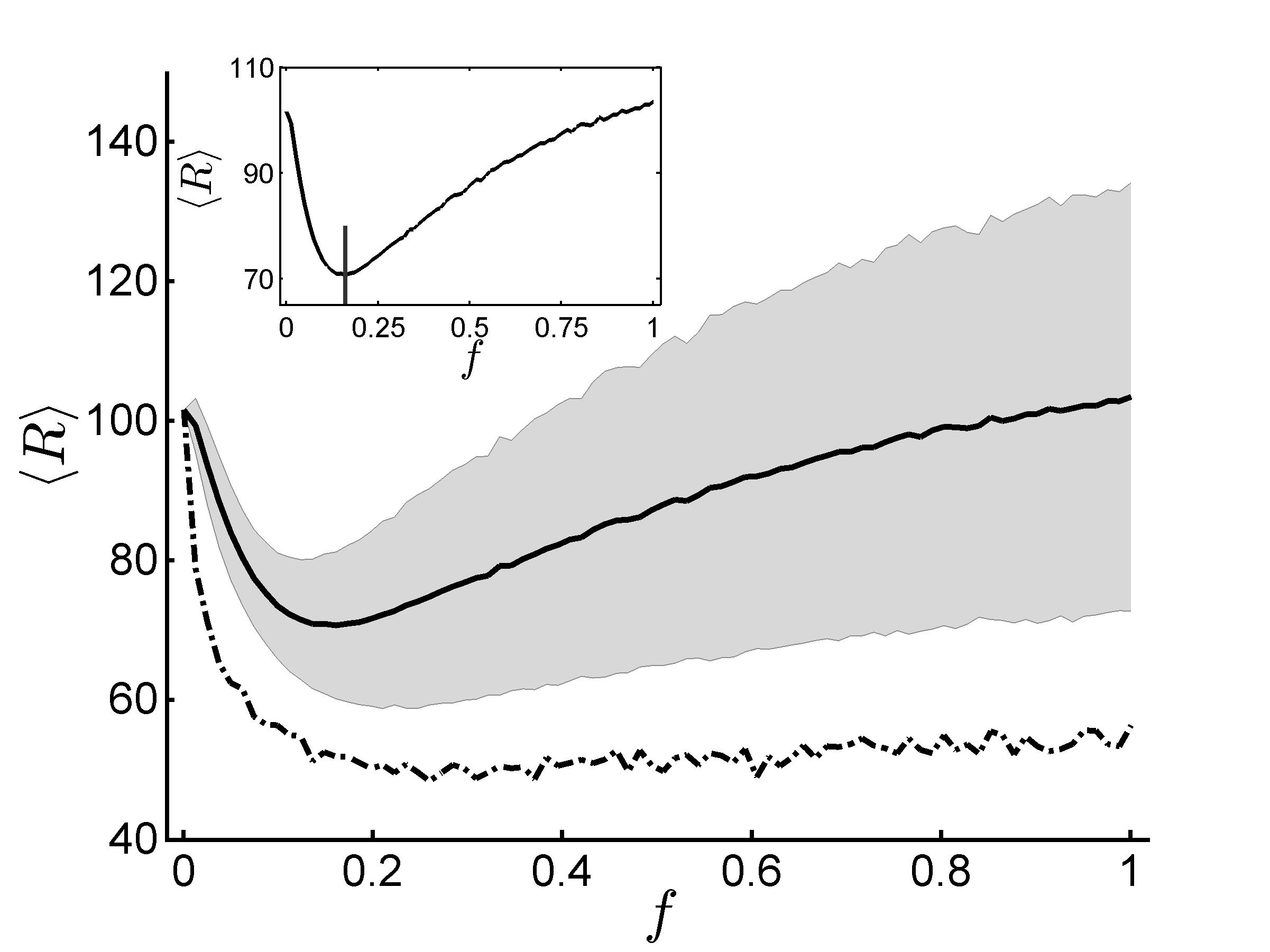}
}
\caption{\label{fig:Figure_3}Relationship of expected synchronizability $\langle R \rangle$ (solid line) with the fraction $f$ of rewired edges of the 3-generation (a) \DSF{} and (b) \PSF{} networks. The shaded areas are representative of the standard deviations (1$\sigma$) of the $R$ values for the ensemble of rewired networks generated for computing $\langle R \rangle$ for any particular value of $f$. The dashed line represents the minimum $R$ value over the ensemble of rewired networks for a given value of $f$. The inset magnifies the $\langle R \rangle$ values, where the vertical line marks the value of $f^{*} = 0.046$ (0.16) for the \DSF{} (\PSF{}) network. Note that we do not rewire ($e$) edges (for a given value of $f$) of the same realization, but generate ensembles of networks with ($e$) rewired edges (for the respective value of $f$). Therefore, one may obtain different values of $f^{*}$ for different realizations, if they were rewired consecutively instead of the procedure as followed here.}
\end{center}
\end{figure}

\begin{figure}
\begin{center}
\includegraphics[height=5.0cm, width=8.5cm]{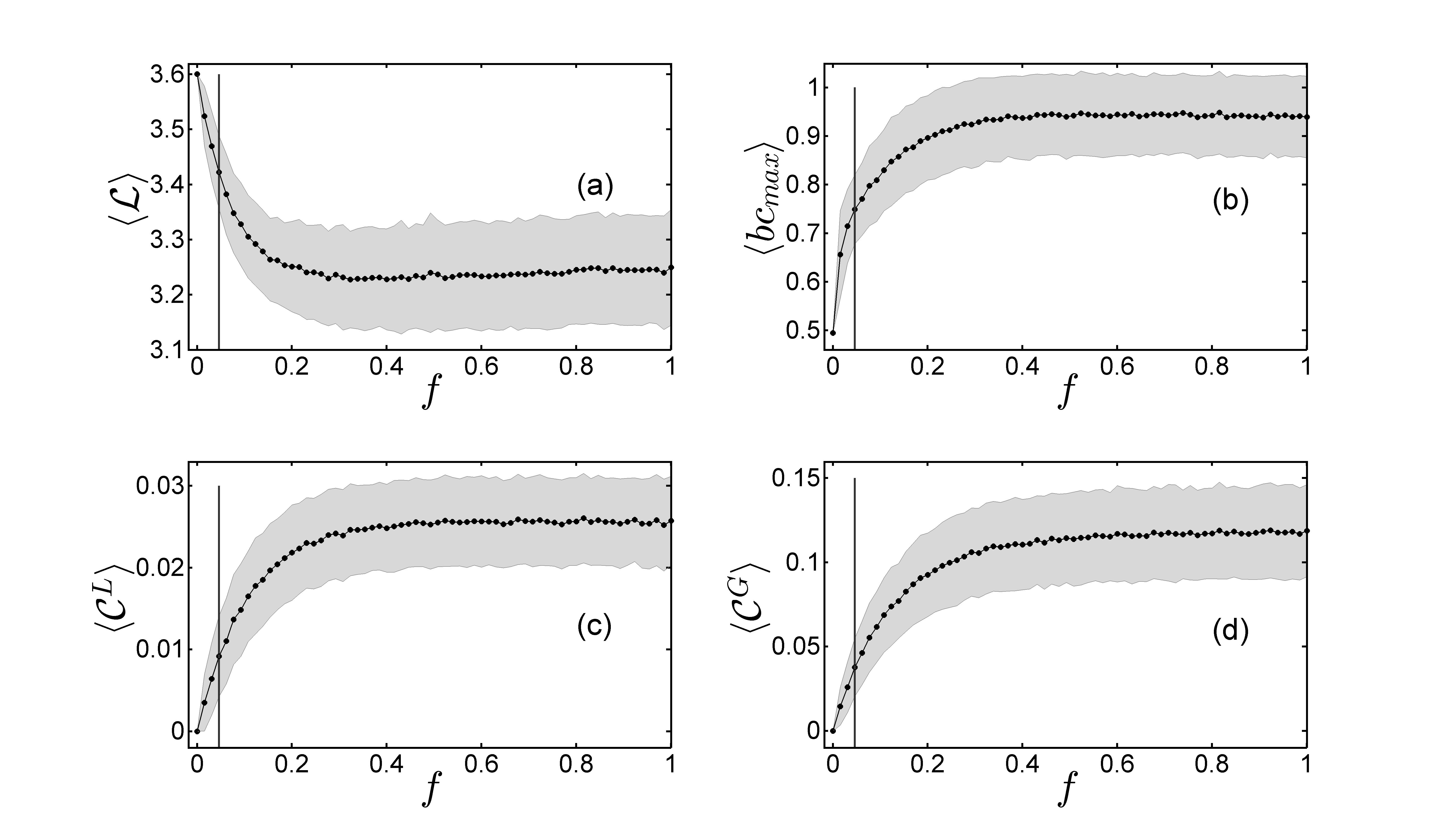}
\caption{\label{fig:Figure_4}Relationship between $f$ and the topological properties (a) $\langle \mathcal{L} \rangle$, (b) $\langle bc_{max} \rangle$, (c) $\langle \mathcal{C}^{L} \rangle$, and (d) $\langle \mathcal{C}^{G} \rangle$ of the associated ensemble of randomly rewired \DSF{} networks. The shaded areas are representative of the standard deviations (1$\sigma$) of the respective topological features of the ensemble of rewired networks (generated for a given value of $f$). The vertical lines indicate the location of $f^{*}$.}
\end{center}
\end{figure}

We consider two paradigmatic network topologies simultaneously exhibiting scale-free degree distributions and hierarchical organization. In the one hand, we study a \DSF{} network developed over 3 generations comprising $N = 81$ nodes and $E = 130$ edges. On the other hand, we investigate a 3-generation \PSF{} network with $N = 123$ nodes and $E = 243$ edges. In both cases, we generate an ensemble of $10^4$ networks by rewiring $e$ (equivalently, a fraction $f = \frac{e}{E}$) pairs of edges of the completely deterministic networks, using the mechanism described in Fig.~\ref{fig:Figure_2}. Further, for a particular value of $f$, we compute the values of $\mathcal{L}$, $bc_{max}$, $\mathcal{C}^{L}$, $\mathcal{C}^{G}$ and $R$ of each network with $e$ randomly rewired links of the ensemble and then estimate the expectation values $\langle \mathcal{L} \rangle$, $\langle bc_{max} \rangle$, $\langle \mathcal{C}^{L} \rangle$, $\langle \mathcal{C}^{G} \rangle$ and $\langle R \rangle$ as the corresponding ensemble means.

We present the variation in the expected synchronizability $\langle R \rangle$ (solid line) with the fraction $f$ of rewired edges of the \DSF{} network in Fig.~\ref{fig:Figure_3}(a). We clearly observe that rewired versions of the otherwise completely \DSF{} network exhibit significantly enhanced, as well as, deteriorated values of synchronizability [Fig.~\ref{fig:Figure_3}(a)]. The dashed line represents the minimum $R$ value over the ensemble of rewired networks for a given value of $f$. The corresponding topologies thus represent approximately `optimally' synchronizable networks for the respective value of $f$. The fluctuations in the minimum $R$ values may be attributed to the relatively small considered ensemble sizes ($10^{4}$), as compared with the much greater variety of possible rewired networks for a given value of $f$. Also, in the inset of Fig.~\ref{fig:Figure_3}(a), we observe a minimal value of $\langle R \rangle$ (highest average synchronizability) for $f$ equal to $f^{*} = 0.046$ (equivalently, 6 rewired edges) of the 81-node network. As $f$ is further increased beyond $f^{*}$, the value of $\langle R \rangle$ increases again, finally saturating at $\langle R \rangle \sim 185$ for $f \gtrsim 0.6$.

Figure~\ref{fig:Figure_3}(b) demonstrates that a similar (and even more pronounced) behaviour of average synchronizability is found in the \PSF{} networks, for which we observe a minimal value of $\langle R \rangle$ for $f^{*} = 0.16$ (equivalently, 39 rewired edges). Moreover, we found similar results (not presented here for brevity) with regard to synchronizability of 4-generation \DSF{} and \PSF{} networks as well.

\begin{figure}
\begin{center}
\includegraphics[height=5.0cm, width=8.6cm]{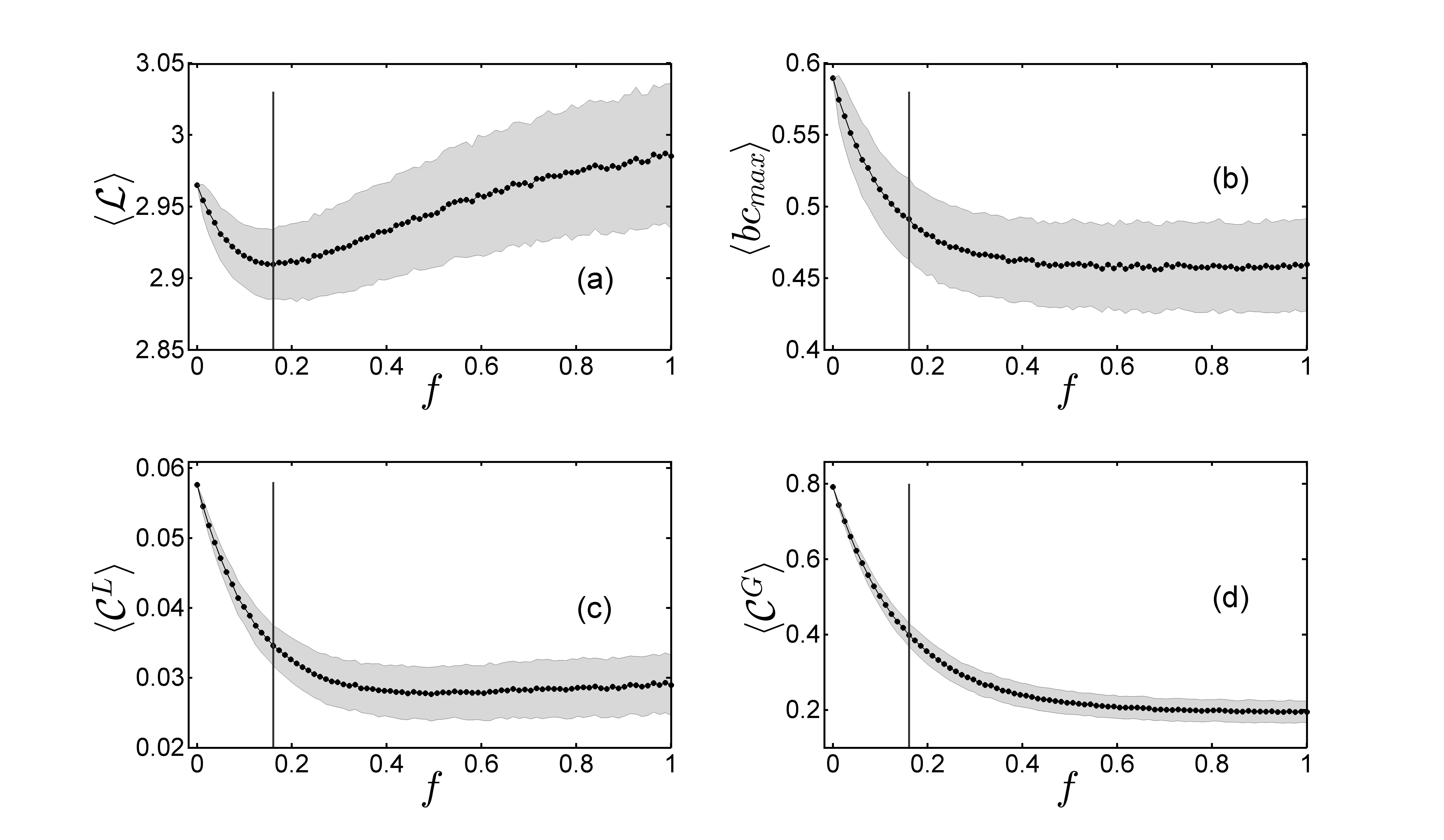}
\caption{\label{fig:Figure_5}Same as in Fig.~\ref{fig:Figure_4} for randomly rewired \PSF{} networks.}
\end{center}
\end{figure}

We further investigate the relationships between $f$ and the topological properties $\langle \mathcal{L} \rangle$, $\langle bc_{max} \rangle$, $\langle \mathcal{C}^{L} \rangle$ and $\langle \mathcal{C}^{G} \rangle$ of the associated ensemble of stochastically rewired \DSF{} networks in Fig.~\ref{fig:Figure_4}. For $f < f^{*}$, the decrease in $\langle \mathcal{L} \rangle$ and the increase in $\langle bc_{max} \rangle$ conform to the decreasing trend of $\langle R \rangle$ (as per the earlier discussion on network properties and their relationship with synchronizability). The value of $\langle \mathcal{C}^{L} \rangle$ (as well as $\langle \mathcal{C}^{G} \rangle$) starts from zero and increases as more edges are rewired. This implies the formation of triangles in the network, which promotes communication between the oscillators, thereby enhancing synchronizability. However, for $f > f^{*}$, further decrease in $\langle \mathcal{L} \rangle$ and increase in $\langle bc_{max} \rangle$ should still improve the average synchronizability, which however only declines from thereon.

Thus, rewiring a few edges ($f < f^{*}$) alters the topological features of the ensemble of networks for better synchronizability. However, when more edges ($f > f^{*}$) are further rewired, it no longer affects on average the topological properties relevant for improving synchronizability, in fact, only undermines it. Hong \etal~\cite{hong2004factors} have previously proposed maximum betweenness centrality as a suitable indicator for predicting synchronizability of networks. They have shown that among various topological factors, such as, short characteristic path length or large heterogeneity of the degree distribution, it is a small value of the maximum betweenness centrality of a network that promotes synchronization~\cite{hong2004factors}. However, this is not corroborated by our results in Fig.~\ref{fig:Figure_4} where we do not observe a strong linear relationship between $\langle R \rangle$ and $\langle bc_{max} \rangle$, as also indicated by a correlation coefficient of 0.776. Similarly, a correlation coefficient of -0.681 rules out a systematic linear dependence between $\langle R \rangle$ and $\langle \mathcal{L} \rangle$. However, a correlation coefficient of 0.847 (0.889) between $\langle R \rangle$ and $\langle \mathcal{C}^{L} \rangle$ ($\langle \mathcal{C}^{G} \rangle$) indicates an appreciable underlying linear relationship. Further, for $f > f^{*}$, the correlation coefficient of 0.939 (0.970) between $\langle R \rangle$ and $\langle \mathcal{C}^{L} \rangle$ ($\langle \mathcal{C}^{G} \rangle$) underlines the above observation.

Analogously to Fig.~\ref{fig:Figure_4}, Fig.~\ref{fig:Figure_5} again shows the relationships between $f$ and the topological properties $\langle \mathcal{L} \rangle$, $\langle bc_{max} \rangle$, $\langle \mathcal{C}^{L} \rangle$ and $\langle \mathcal{C}^{G} \rangle$ of the associated ensemble of rewired \PSF{} networks. In this case, we observe a clear relationship between $\langle R \rangle$ and $\langle \mathcal{L} \rangle$, further corroborated by a correlation coefficient of 0.987. On the other hand, a possible linear relationship between $\langle R \rangle$ and $\langle bc_{max} \rangle$, $\langle \mathcal{C}^{L} \rangle$ and $\langle \mathcal{C}^{G} \rangle$ is ruled out by correlation coefficients of -0.25, -0.175 and -0.373, respectively.

Taken together, we notice that the topological features of the ensembles of rewired \DSF{} (Fig.~\ref{fig:Figure_4}) and \PSF{} (Fig.~\ref{fig:Figure_5}) networks exhibit certain contrasting variations, as $f$ is tuned from 0 to 1. Prior to saturation, the $bc_{max}$ of the rewired \DSF{} networks (Fig.~\ref{fig:Figure_4}(b)) initially increases with $f$, as opposed to a corresponding decrease in $bc_{max}$ observed for the rewired \PSF{} networks (Fig.~\ref{fig:Figure_5}(b)). On the contrary, both clustering coefficients $\langle \mathcal{C}^{L} \rangle$ and $\langle \mathcal{C}^{G} \rangle$ increase with $f$ until saturation for rewired \DSF{} networks (Fig.~\ref{fig:Figure_4}(c, d)), which however display a decreasing trend in the case of rewired \PSF{} networks (Fig.~\ref{fig:Figure_5}(c, d)).

We now compare the synchronizability of rewired \DSF{} and \PSF{} networks with that of random scale-free networks generated using the classical BA model of growth and preferential attachment~\cite{barabasi1999emergence}. In this regard, we consider an ensemble of 100 such random scale-free networks of 81 nodes (123 nodes) each for comparison with rewired \DSF{} (\PSF{}) networks, respectively. While generating the BA networks, we incorporate the growing character of the network by starting with a small number of vertices and at every time step introducing a new vertex and linking it to 2 vertices already present in the system, until the network comprises 81 (123) nodes. Preferential attachment is incorporated by assuming that the probability $\Pi_{i}$ that a new node will be connected to node $i$ depends on the degree $k_{i}$ of node $i$, such that $\Pi_{i} = \frac{k_{i}}{\sum\limits_{j} k_{j}}$. The 81-node (123-node) BA networks have a total of 158 (242) edges in each realization. The $\langle R \rangle$ values of the considered ensemble of 81-node (123-node) BA networks turn out to be 36.74 (49.75), which is much smaller than the minimum $R$ values among the ensembles of rewired \DSF{} (\PSF{}) networks for different $f$, presented in Fig.~\ref{fig:Figure_3}. Thus, random scale-free networks generated using the classical BA model appear to promote synchronizability better than randomly rewired \DSF{}, as well as, \PSF{} networks. We outline further investigations to unveiling the reasons for this behaviour as a subject of future research.


\section{\label{sec:Conclusion}Conclusion}

Many real-world complex networks simultaneously exhibit generic features of scale-free topology along with hierarchical organization. In this regard, two notable models which simultaneously capture the two different topological properties are the deterministic and pseudofractal scale-free networks. These models comprise completely deterministic processes underlying the formation of the respective networks. However, real-world networks are presumably neither completely deterministic, nor perfectly hierarchical. Thus, a practical model of such networks should feature an aspect of randomness, while exhibiting scale-free and hierarchical design. For this purpose, we suggested preserving the scale-free degree distribution of the deterministic networks we start with, while tweaking the hierarchical structure by rewiring them. Specifically, we hypothesized that perfectly hierarchical scale-free networks (generated by the deterministic rules of the aforementioned models) with randomly rewired links may provide more realistic representatives of associated real-world topologies than perfectly hierarchical ones.

The desired operational state in many complex systems often concurs with the synchronized motion of dynamical units coupled on a networked architecture. Consequently, we utilized the analytical framework of master stability function (\MSF{}) in investigating synchronizability of dynamical systems coupled on the proposed network structures. Interestingly, this revealed that the process of rewiring is capable of significantly enhancing, as well as, deteriorating the synchronizability of the resulting networks. Importantly, when a certain critical fraction of edges of the otherwise completely deterministic networks were rewired, it optimized the average synchronizability of the resulting topologies. This observation is, however, different from \emph{Braess's paradox} where the \emph{addition} of edges undermines synchrony in complex oscillator networks~\cite{witthaut2012braess}. We also investigated the influence of rewiring links on some key topological properties (average path length, maximum betweenness centrality, average local clustering coefficient and global clustering coefficient) of the resulting networks and, in turn, their relation to the synchronizability of the associated topologies demonstrating distinct behaviours in these different models of hierarchical scale-free networks. We speculate that an interplay between the various topological properties of the networks, in particular, their average path lengths and clustering coefficients in a trade-off lead to an `optimal' value of synchronizability when rewiring the respective networks.

In a related context, we recall that networks exhibiting the small-world property have been considered conducive for synchronization~\cite{nishikawa2003heterogeneity, menck2013basin}. However, \MSF{}-based measurements of the synchronizability of Watts-Strogatz networks did not reveal exclusive features in the small-world regime~\cite{hong2002synchronization}. Importantly, the critical fraction of rewired edges (for maximal synchronizability) in the hierarchical scale-free networks considered here, roughly corresponds to a similar value for typical Watt-Strogatz networks to exhibit small-world behaviour. Specifically, we also found that rewiring a few edges of the deterministic scale-free, as well as, pseudofractal scale-free networks generated a topology with significantly enhanced or `optimal' synchronizability, which did not exhibit major improvements thereafter, as the fraction of rewired edges was further increased.

The aforementioned results may have potential implications in the design of complex networks (simultaneously exhibiting hierarchical structure and scale-free behaviour) for better synchronizability. A more challenging problem is that of comparing real-world topologies with rewired versions of deterministic scale-free hierarchical networks explored here, in ascertaining a possible deterministic backbone of certain practical networks and the proportion of randomness in the same. Any efforts in this direction could certainly provide deeper insights into the developmental processes and synchronizability of many practical networked dynamical systems simultaneously displaying hierarchical structure and scale-free behaviour.


\acknowledgments
CM and RVD have been supported by the German Federal Ministry of Education and Research (BMBF) via the Young Investigators Group CoSy-CC\textsuperscript{2} (grant no.\ 01LN1306A). JK \& RVD acknowledge support from the IRTG 1740/TRP 2011/50151-0, funded by the DFG/FAPESP. The authors gratefully acknowledge the European Regional Development Fund (ERDF), the German Federal Ministry of Education and Research (BMBF) and the Land Brandenburg for supporting this project by providing resources on the high performance computer system at the Potsdam Institute for Climate Impact Research.


\bibliographystyle{eplbib}
\bibliography{References}

\end{document}